# Metal-Insulator Transition in Variably Doped (Bi$_{1-x}$Sb$_x$)$_2$Se$_3$ Nanosheets

Chee Huei Lee,[†,‡] Rui He,[§,*] ZhenHua Wang,[†,¶] Richard L.J. Qiu,[†] Ajay Kumar,[‡] Conor Delaney,[§] Ben Beck,[§] T. E. Kidd,[§] C. C. Chancey,[§] R. Mohan Sankaran,[‡] and Xuan P. A. Gao,[†,*]

[†]Department of Physics, Case Western Reserve University, Cleveland, OH 44106, U.S.A.

[‡]Department of Chemical Engineering, Case Western Reserve University, Cleveland, OH 44106, U.S.A.

[§]Department of Physics, University of Northern Iowa, Cedar Falls, IA 50614, U.S.A.

[¶]Shenyang National Laboratory for Materials Science, Institute of Metal Research, and International Centre for Materials Physics, Chinese Academy of Sciences, Shenyang 110016, People's Republic of China

*Email: (R.H.) rui.he@uni.edu; (X.P.A.G.) xuan.gao@case.edu

**Abstract**

Topological insulators are novel quantum materials with metallic surface transport, but insulating bulk behavior. Often, topological insulators are dominated by bulk contributions due to defect induced bulk carriers, making it difficult to isolate the more interesting surface transport characteristics. Here, we report the synthesis and characterization of nanosheets of topological insulator Bi$_2$Se$_3$ with variable Sb-doping level to control the electron carrier density and surface transport behavior. (Bi$_{1-x}$Sb$_x$)$_2$Se$_3$ thin films of thickness less than 10 nm are prepared by epitaxial growth on mica substrates in a vapor transport setup. The introduction of Sb in Bi$_2$Se$_3$ effectively suppresses the room temperature electron density from ~4×10$^{13}$/cm$^2$ in pure Bi$_2$Se$_3$ ($x$ = 0) to ~2×10$^{12}$/cm$^2$ in (Bi$_{1-x}$Sb$_x$)$_2$Se$_3$ at $x$ ~0.15, while maintaining the metallic transport behavior. At $x \gtrsim$ ~0.20, a metal-insulator transition (MIT) is observed indicating that the system has transformed into an insulator in which the metallic surface conduction is blocked. In agreement with the observed MIT, Raman spectroscopy reveals the emergence of vibrational modes arising from Sb-Sb and Sb-Se bonds at high Sb concentrations, confirming the appearance of $Sb_2Se_3$ crystal structure in the sample. These results suggest that nanostructured chalcogenide films with controlled doping can be a tunable platform for fundamental studies and electronic applications of topological insulator systems.

1. **Introduction**

Three dimensional (3D) topological insulators (TIs) have attracted significant research interest due to the presence of spin-momentum-locked metallic surface states in the insulating bulk band gap.[1-3] This new type of quantum behavior provides rich opportunities for studying fundamental physics,[4-6] as well as practical applications.[7] However, achieving clear experimental signatures for TI properties can be difficult. In particular, accessing metallic Dirac surface states in TIs is challenging since bulk carriers often dominate and overwhelm the surface contribution. Exposing TIs to air has been shown to generate additional bulk carriers,[8,9] adding another hurdle to the inspection of surface states.

Recent attention has been paid to the realization and study of phase transitions between a TI phase and a non-topological metal[10,11] or band insulator phase.[12] Investigations of such phase transitions should shed light on the physics of topological phase transitions, critical for device applications of these exotic quantum materials. Currently, there are only a few reports of topological to non-topological phase transitions, including those in bulk $TlBi(S_{1-x}Se_x)_2$[10] and in thin films of $(Bi_{1-x}In_x)_2Se_3$ grown by molecular beam epitaxy (MBE).[12]

Among the binary compounds of Bi, Sb, Se, and Te, three ($Bi_2Se_3$, $Bi_2Te_3$ and $Sb_2Te_3$) have been identified as 3D TIs, but $Sb_2Se_3$ is a trivial band insulator with a bandgap of ~ 1.2 eV.[2] $Bi_2Se_3$ is one of the most widely studied 3D TI materials due to its bulk band gap of 0.3 eV, which makes it a promising material for room temperature applications.[2, 7] While nanowires, nanoribbons,[13, 14] and MBE-grown thin films[15-18] of $Bi_2Se_3$ have been synthesized to exploit the large surface-to-volume ratio of these nanostructures, the residual electron concentration is still high in these 'intrinsic' samples and the metallic temperature-dependent resistivity is overwhelmed by the bulk contribution. Elemental doping of $Bi_2Se_3$ and $Bi_2Te_3$ bulk crystals or nanoribbons[19-22] to obtain ternary or quaternary compounds,[23, 24] and electrical gating[25-28] techniques have been used to modulate the Fermi level and suppress bulk carriers. Ternary $(Bi_{1-x}Sb_x)_2Te_3$ films have been grown by MBE and found to host topological surface states over the whole range of $x$ ($0 < x < 1$).[29]

Exploiting the fact that $In_2Se_3$ is a band insulator, Brahlek *et al*. grew $(Bi_{1-x}In_x)_2Se_3$ films over a range of values of $x$ using MBE and were able to demonstrate a transition from TI to band insulator at $x \sim 0.25$.[12] Here, we show precise control of electronic doping and the metal to insulator transition in $(Bi_{1-x}Sb_x)_2Se_3$ films for the first time. Large-area (~2-3 $cm^2$) nanosheets of $(Bi_{1-x}Sb_x)_2Se_3$ with a wide-range of Sb concentrations were synthesized *via* van der Waals epitaxial growth on mica substrates, as recently reported by Peng *et al*. for pure $Bi_2Se_3$.[30] As the Sb concentration, $x$, is varied from 0 to ~0.26, bulk carrier densities are effectively reduced by more than an order of magnitude from ~$4\times10^{13}/cm^2$ at $x = 0$ to ~$2\times10^{12}/cm^2$ when the doping level $x>0.15$. When the Sb fraction is increased further ($x>\sim0.20$), a MIT is revealed in the temperature dependent electrical transport data due to a transition from a TI with metallic surface states to an insulator state without metallic surface conduction, as expected for pure $Sb_2Se_3$.[2]

Raman spectroscopy confirms that new peaks arise from Sb-Sb and Sb-Se bonds when the doping level of Sb is between 10% and 18%, revealing the formation of $Sb_2Se_3$ crystalline structures in the $Bi_2Se_3$ host lattice. The evolution of Raman spectra as a function of Sb doping concentration is understandable in the context of a rhombohedral/orthorhombic mixed phase formed in the $(Bi_{1-x}Sb_x)_2Se_3$ system. The electronic properties are explained by a percolation mechanism.

2. **Experimental Techniques**

2.1 $(Bi_{1-x}Sb_x)_2Se_3$ nanosheet synthesis

Figure 1 illustrates a schematic diagram of the experimental setup used to synthesize nanosheets of $(Bi_{1-x}Sb_x)_2Se_3$ on mica substrates. In a typical experiment, solid powder sources of $Bi_2Se_3$ and $Sb_2Se_3$ (99.999%, Alfa Aesar) were loaded into a tube furnace (Lindberg/Blue M) by quartz transfer rods. The $Sb_2Se_3$ powder was placed upstream of the furnace in order to keep the concentration low both in the vapor and in the as-deposited thin film. A freshly-cleaved mica substrate (SPI supplies, Grade V-5 Muscovite, 5cm by 1cm) was placed at 13-18 cm downstream from the center of the furnace where the growth temperature at that region was between 280-380 °C. The furnace was initially pumped down to 20 mTorr, then a carrier gas of Ar with 10% $H_2$ at a flow rate of 1000 sccm was used to transport the vapor from the powder sources to the substrate. The furnace was heated to 500 °C within 10 min, at a pressure of 100 Torr. After the growth was carried out for 30 min, both transfer rods were pulled out from the furnace to terminate the growth. The furnace was then allowed to cool down naturally to room temperature with the carrier gas flowing. The level of the Sb dopant concentration in the $(Bi_{1-x}Sb_x)_2Se_3$ films was controlled by varying the position of the $Sb_2Se_3$ powder source in the furnace (which corresponds to different source temperatures and therefore to different vapor concentrations of Sb$_2$Se$_3$). In our experiments, the Bi$_2$Se$_3$ powder was placed at the center of furnace and the Sb$_2$Se$_3$ powder was placed 11.5 to 13 cm upstream from the center of the furnace, a location corresponding to a temperature of ~445 to 460 °C when the furnace is set to 500 °C. The composition of the nanosheets was examined by X-ray photoelectron spectroscopy (XPS) and EDS (energy dispersive X-ray) equipped in a high resolution TEM (transmission electron microscope).

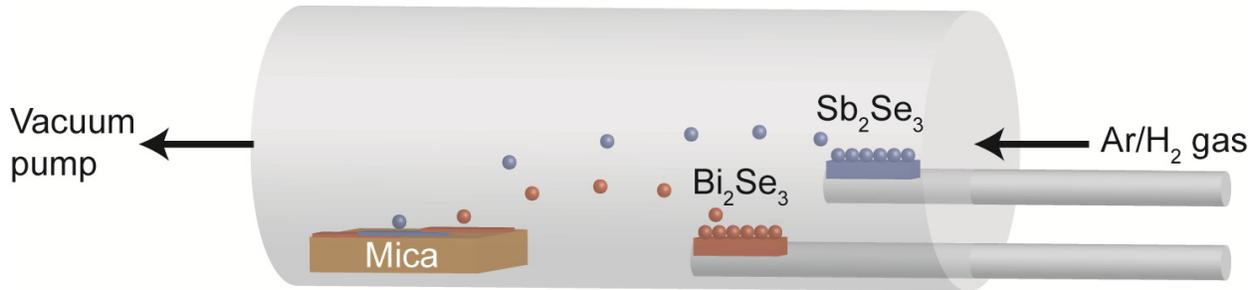

**Fig. 1**. Schematic diagram of the vapor transport setup used for growth of Sb-doped Bi$_2$Se$_3$ nanosheets *via* van der Waals epitaxy on mica substrates.

2.2 Electrical transport characterization

To perform electrical transport characterization, the Sb-doped Bi$_2$Se$_3$ samples were first cut to a rectangular shape (2 × 10 mm). A Hall bar geometry was fabricated using indium as electrical contacts. Because of the large area of our thin films, photolithographic steps were completely avoided since these involve organic chemicals that could potentially contaminate or oxidize the films. Electrical measurements were conducted by a lock-in technique at low-frequency using a Quantum Design Physical Properties Measurement System cryostat. A typical source-drain bias was 50 mV. Hall effect measurements were made by sweeping the magnetic field up to ±9 T and recording the Hall resistance, $R_{xy}$. All data were anti-symmetrized to remove longitudinal components. The low field Hall coefficient, $R_H$, was obtained from a linear fitting of $R_{xy}$ between ±1 T. Similarly to literature reports on other TI materials,[8] we observed an aging effect when samples are exposed to air for a long time. Therefore, unless specially noted, the electrical measurements were carried out within 1 to 2 hrs after the samples were grown to minimize environmental doping.

### 2.3 Raman spectroscopy

Raman scattering is a sensitive, high-resolution probe of crystal structure and the chemical composition of materials. With a tightly focused laser spot, we can study local properties of the nanosheets. $Bi_2Se_3$ crystallizes into a rhombohedral structure with quintuple layers ordered in a $Se_1$-Bi-$Se_2$-Bi-$Se_1$ sequence and is a TI.[2] In contrast, $Sb_2Se_3$ is not a TI and has a completely different orthorhombic crystalline structure with four $Sb_2Se_3$ molecules in a unit cell, easily cleaved along the (010) plane.[31] Changes in crystalline microstructures and chemical compositions of materials can be revealed by the changes in the positions and/or bandwidths of the peaks in the Raman spectra. In this study, Raman spectra were obtained at room temperature using a Horiba LabRam Raman Microscope system. The excitation laser light was 532 nm, and a 100× objective lens was used to focus the laser light to a spot of less than 0.5 μm. Because the TI material $Bi_2Se_3$ and the substrate mica are not good conductors of heat, a low laser power of about 15 μW was used during the Raman measurements to avoid excessive heating of the nanosheets.

### 3. Results and Discussion

Optical images of $(Bi_{1-x}Sb_x)_2Se_3$ nanosheets with $x = 0$ and $0.18$ are shown in Fig. 2A, B. The image for a pure $Bi_2Se_3$ nanosheet sample ($x = 0$, Fig. 2A) displays a continuous film with triangular or hexagonal flakes approximately tens of microns in size along a surface. As Sb was introduced, the flakes lose their triangular/hexagonal shape and become irregularly shaped. This morphology change is likely a result of the formation of mixed rhombohedral and orthorhombic crystal structures of pure $Bi_2Se_3$ (rhombohedral) and pure $Sb_2Se_3$ (orthorhombic), consistent with the bulk $(Bi_{1-x}Sb_x)_2Se_3$ solution.[23, 32] This formation of mixed phases is revealed by Raman spectra (see Fig. 3). To obtain the thickness of the synthesized nanosheets, the films were gently scratched and imaged by atomic force microscopy (AFM) (see Fig. 2C for pure $Bi_2Se_3$ sample and Fig. 2D for $(Bi_{1-x}Sb_x)_2Se_3$ sample with $x\sim0.10$). For growth time of 30 min, the thickness of the $(Bi_{1-x}Sb_x)_2Se_3$ nanosheets ranged from 6 to 10 nm. We did not observe a large variation in thickness between pure $Bi_2Se_3$ and doped samples. The incremental change in the thickness of the sample's height profile as obtained by AFM line scans was ~1 nm (Fig. 2E), which corresponds to the thickness of 1 quintuple layer (QL) (1 QL ≈ 0.96 nm thick),[33] indicating an epitaxial growth mechanism. We further evaluated the crystalline structure of the $(Bi_{1-x}Sb_x)_2Se_3$ nanosheets by TEM. Fig. 2F shows a low magnification bright field TEM image of the pure $Bi_2Se_3$ film. A high resolution TEM image shown in Fig. 2G indicates that the film is crystalline. The lattice spacing between the $(11\bar{2}0)$ planes was measured to be ~0.2 nm, consistent with the lattice parameter of $Bi_2Se_3$.[34, 35] In Fig. 2H, the corresponding fast Fourier transform (FFT) pattern shows 6-fold symmetry [0001], in agreement with the hexagonal structure of $Bi_2Se_3$ along its c-axis. We note that Sb-doped $Bi_2Se_3$ nanosheets also show crystalline structures but different microstructures and vacancies are easily found in TEM imaging (dashed lines in Fig. 2I for a $(Bi_{1-x}Sb_x)_2Se_3$ nanosheet with $x = 0.10$ ).

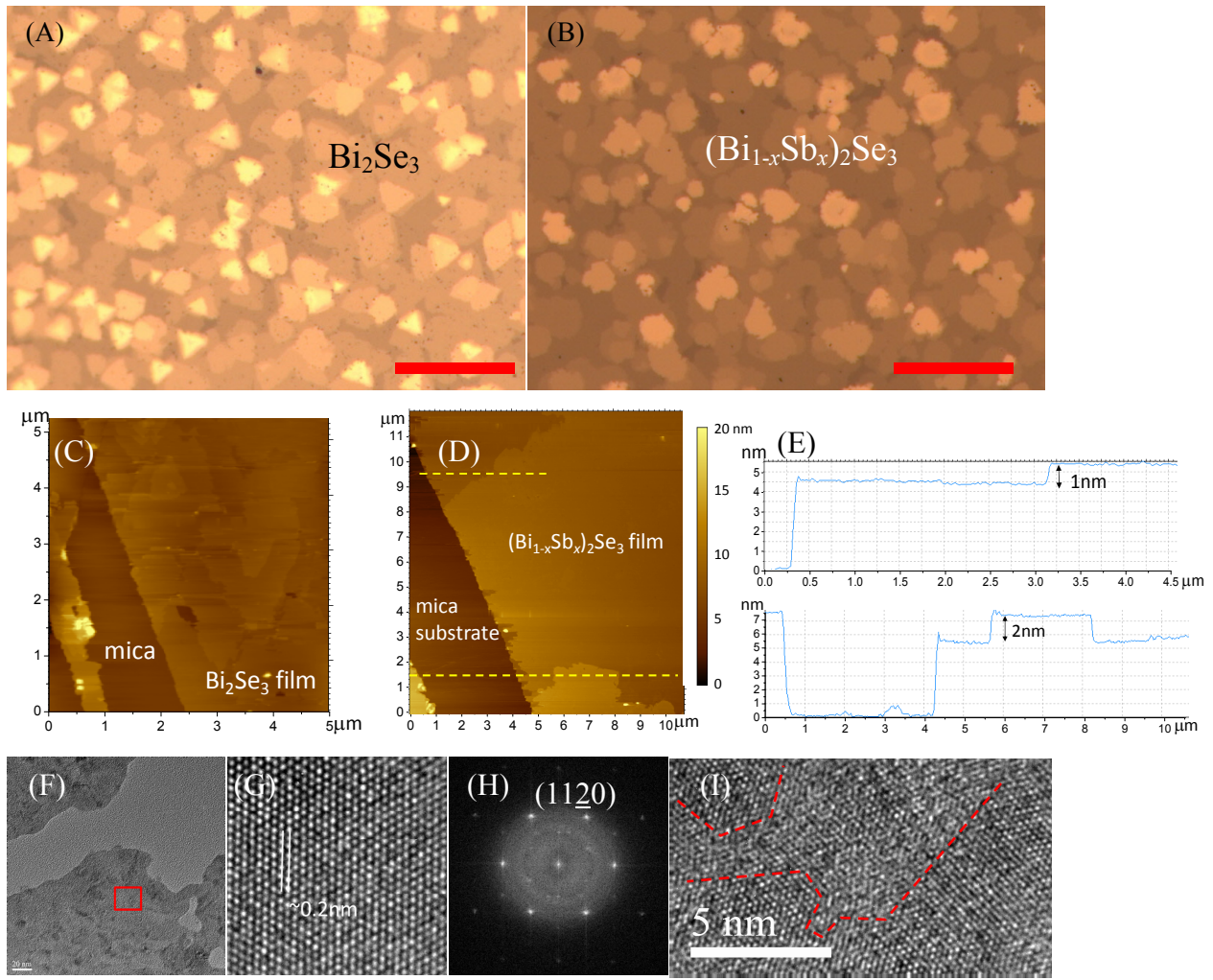

**Fig. 2.** (A, B) Optical images of pure ($x = 0$) and doped $(Bi_{1-x}Sb_x)_2Se_3$ nanosheets ($x \approx 0.18$) grown on mica, respectively. Scale bars are 100 μm. (C, D) AFM images of pure $Bi_2Se_3$ and $(Bi_{1-x}Sb_x)_2Se_3$ ($x \approx 0.10$) thin films. A physical scratch was intentionally made to reveal the mica surface and determine the film thickness. (E) Height profiles along the dashed lines indicated in panel (D). (F) TEM image of pure $Bi_2Se_3$ thin film. (G) HRTEM image of pure $Bi_2Se_3$ thin film, taken along the $Bi_2Se_3$ [0001] direction, showing the crystalline structure. The lattice spacing is measured to be ~0.2 nm, consistent with the spacing of $(11\bar{2}0)$ planes in $Bi_2Se_3$ bulk material.[34, 35] (H) Fast Fourier transform (FFT) of the HRTEM image shown in (G). (I) HRTEM image of $(Bi_{1-x}Sb_x)_2Se_3$ ($x \approx 0.10$). The dashed lines show the boundary between crystalline grains.

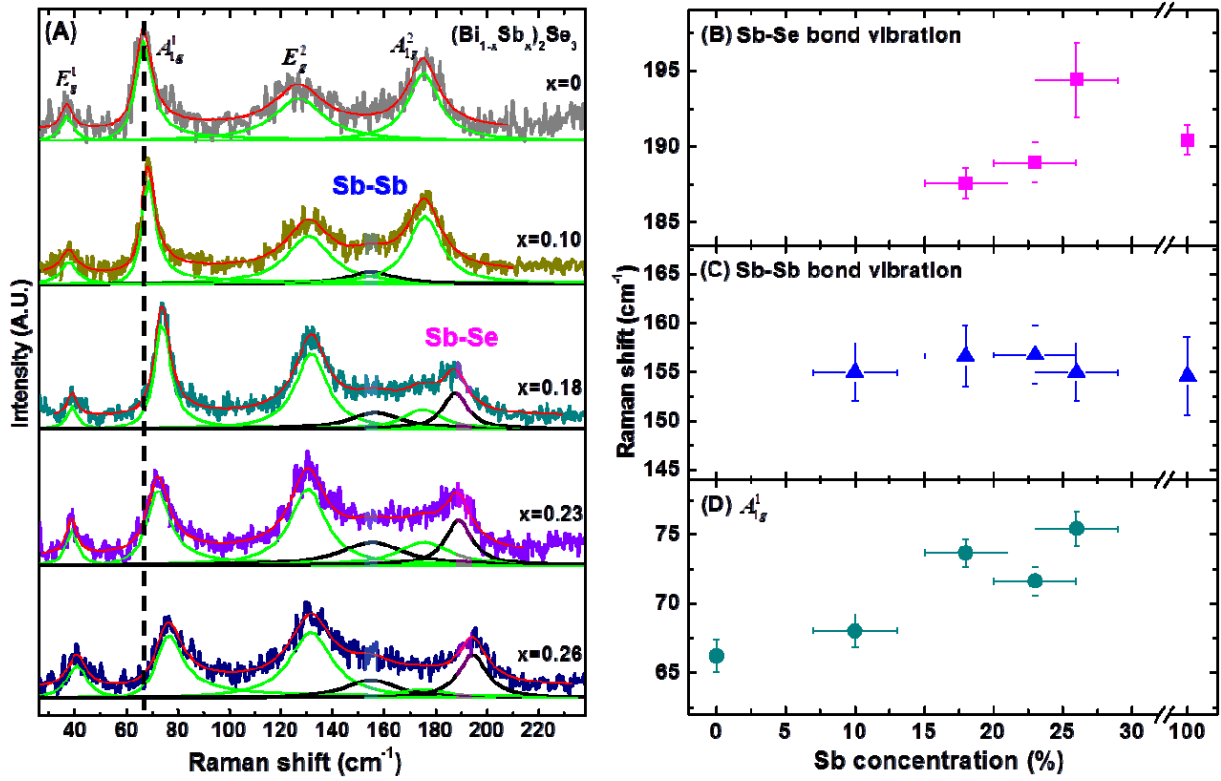

**Fig. 3.** (A) Typical Raman spectra from $(Bi_{1-x}Sb_x)_2Se_3$ nanosheets with different doping levels. The thicknesses of the probed nanosheets are $5 \pm 2.5$ nm and are determined by AFM. Each spectrum is fit with Lorentzian/Fano lineshapes, and the red lines are overall profiles from the fittings. The dashed vertical line highlights the position of the $A_{1g}^1$ mode from pure $Bi_2Se_3$ nanosheets. The blue and pink vertical bars highlight the positions of the Sb-Sb bond and Sb-Se bond vibrations, respectively. (B-D) Positions of the three Raman peaks (highlighted in panel (A)) as a function of doping level. The Sb-Sb and Sb-Se bond vibrational frequencies for pure $Sb_2Se_3$ (Sb concentration 100%) in panels (B) and (C) were obtained from Ref. [38].

Fig. 3A shows typical Raman spectra obtained from nanosheets of $(Bi_{1-x}Sb_x)_2Se_3$ with different $x$ values. For the pure $Bi_2Se_3$ nanosheets ($x = 0$), four Raman modes with $E_g$ and $A_{1g}$ symmetries are observed. The two $E_g$ modes are from in-plane vibrations of Bi-Se$_1$ pairs, and the two $A_{1g}$ modes are due to out-of-plane vibrations of Bi-Se$_1$ pairs.[36, 37] As $x$ is increased to 0.10, the spectrum is similar to that obtained for the $x = 0$ sample, except for the emergence of a weak mode centered around 155 cm$^{-1}$. This mode is present in the Raman spectrum from Sb$_2$Se$_3$ crystals and is associated with the Sb-Sb bond vibrations.[38] As the doping level $x$ continues to increase to 0.18 or above, this mode from Sb-Sb bonds remains in the vicinity of 155 cm$^{-1}$ (highlighted by the vertical blue bar). The position of this Raman peak from different samples is summarized in Fig. 3C.

The most prominent change in the Raman spectra when $x$ is increased to 0.18 is the appearance of a new mode at about 190 cm$^{-1}$ (highlighted by the pink vertical bar in Fig. 3A and summarized in Fig. 3B). This mode has been attributed to Sb-Se bond vibrations in Sb$_2$Se$_3$ crystals.[38] This observation indicates that the Bi$_2$Se$_3$ host lattice (rhombohedral phase) undergoes a major transformation by incorporating Sb$_2$Se$_3$ guest lattice structures (orthorhombic phase) as $x$ approaches 0.18, as revealed by the presence of both Sb-Sb and Sb-Se bond vibrations in the Raman spectra at this doping level. Note that the intensity of the Sb-Se bond vibration is negligible when the doping level is relatively low ($x \sim 0.10$). In this case, the concentration of the orthorhombic Sb$_2$Se$_3$ phase in the Bi$_2$Se$_3$ host lattice is low. Independent study by electrical transport (discussed below) shows significant carrier density suppression and a MIT at a similar doping level ($x \sim 0.20$). Hence, the incorporation of Sb$_2$Se$_3$ lattice structure is linked to the reduction of bulk carrier density and could be correlated with the topological transition in (Bi$_{1-x}$Sb$_x$)$_2$Se$_3$.

Figure 3D shows the frequency of the $A_{1g}^1$ out-of-plane mode (highlighted by the dashed vertical line in Fig. 3A), which arises from the vibration of the Bi-Se$_1$ pairs in phase in the Bi$_2$Se$_3$ host lattice, as a function of doping concentration. The frequency of this mode remains essentially unchanged as $x$ changes from 0 to 0.10. However, this $A_{1g}^1$ mode displays a blueshift by about 5 cm$^{-1}$ when $x$ increases to 0.18, though there is no apparent change in its frequency for $x$ between 0.18 and 0.26. Because this mode originates from the $Bi_2Se_3$ rhombohedral phase, our Raman data suggest that the rhombohedral crystalline structure is largely preserved when the Sb concentration is low ($x$ around 0.10). As the doping level $x$ increases to about 0.18 or above, a rhombohedral/orthorhombic mixed phase is formed in the doped system as shown by the presence of Raman modes from both the rhombohedral ($Bi_2Se_3$) phase ($A_{1g}^1$ peak) and the orthorhombic ($Sb_2Se_3$) phase (Sb-Sb and Sb-Se bond vibrations).

To illustrate the effects of substituting Bi with Sb on the electronic properties of topological insulator $Bi_2Se_3$, we performed temperature and magnetic field dependent transport measurement of $(Bi_{1-x}Sb_x)_2Se_3$ nanosheet samples with variable $x$. In our Hall measurement, a negative Hall coefficient was found, indicating electron carriers. Similar to prior work on Sb doped $Bi_2Se_3$ nanoribbons,[21] we found that a few percent Sb doping drastically reduces the electron density $n_e$. The doping concentration-dependent low-field Hall coefficients, $R_H$, and extracted carrier density $n_e = -1/(eR_H)$ are plotted in Fig. 4A and B for $T$ = 300 K. Noticeably, most of the Sb doping induced drop in $n_e$ happens between $x$ = 0 and $x$ = 0.10, before the Sb-Sb bond vibration starts to appear in the Raman spectra. When the Sb fraction is increased further, the resistivity of the sample increases more rapidly and a transition from metallic temperature-dependent resistance to insulating behavior was observed, indicating a MIT between 15 and 22% Sb doping as shown in Fig. 4C.

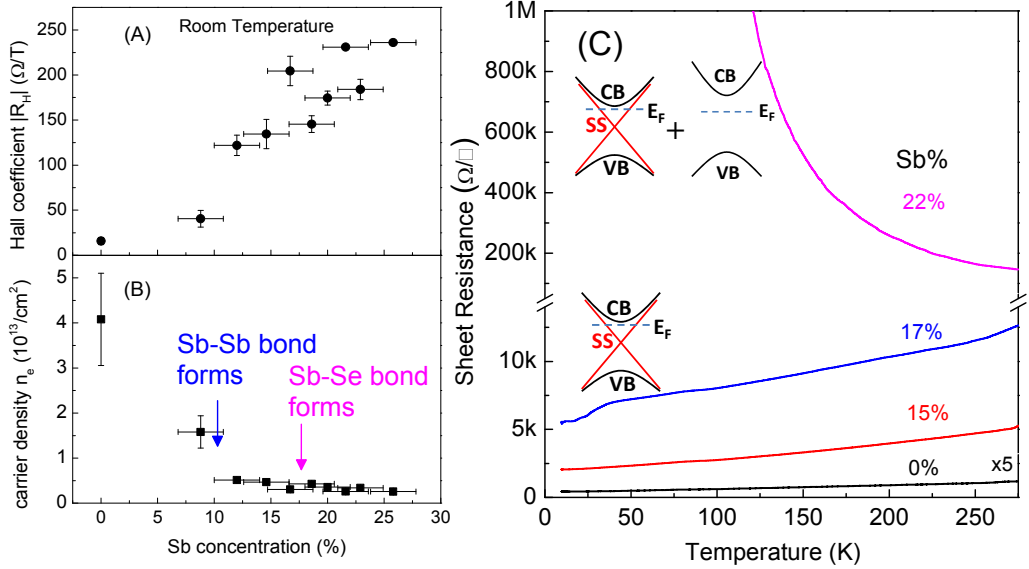

**Fig. 4.** (A) The absolute value of low-field Hall coefficient, |$R_H$|, vs. the Sb concentration in Sb-doped $Bi_2Se_3$ nanosheets. (B) Sheet carrier density vs. Sb concentration in Sb-doped $Bi_2Se_3$ nanosheets. The blue and pink arrows mark the corresponding Sb% where the Sb-Sb and Sb-Se bond vibrations appear in the Raman spectra. (C) Sheet resistance as a function of temperature for pure $Bi_2Se_3$ nanosheets and Sb-doped $Bi_2Se_3$ nanosheets with varying dopant levels, showing a metal-insulator transition around 20% of Sb doping. The insets illustrate the appearance of topologically trivial phase without conducting surface states in band gap across the metal-insulator transition. CB, VB, SS, and $E_F$ represent conduction band, valence band, surface state, and Fermi level, respectively.

We carefully examined the temperature dependence of the sheet resistance for the pure and doped $Bi_2Se_3$ samples. The overall data are presented in Fig. 4C. For the pure $Bi_2Se_3$ sample, the typical sheet resistance, $R_\square$, is ~250 $\Omega\square^{-1}$ at room temperature, in excellent agreement with previous reports for $Bi_2Se_3$.[30] The Hall coefficient, $R_H$, is -12.9 $\Omega$/T, which corresponds to an electron density, $n_e$, of $4.8\times10^{13}$ cm$^{-2}$ (equivalent volume density ~$4.8\times10^{19}$cm$^{-3}$). The temperature dependence of the sheet resistance, $R_\square(T)$, shows a typical metallic behavior as the resistance increases with temperature. Note that at such high sheet carrier densities, the Fermi energy is located above the conduction band minima and bulk carriers contribute to most of the metallic conductivity of samples.[16, 18] The Hall mobility, $\mu_H = \sigma|R_H|$, is deduced to be ~573 cm$^2$/Vs at 300 K. The $\mu_H$ at 10 K is estimated to be ~1450 cm$^2$/Vs ($R_\square$(10K)~88 $\Omega\square^{-1}$). This value is similar to that found in previous studies,[30] implying that our samples are of comparable quality to large area Bi$_2$Se$_3$ thin films grown by others. For the doped Bi$_2$Se$_3$ samples, $R_H$ was measured to be ~ -134 $\Omega$/T at a Sb concentration of 15%. The corresponding carrier density, $n_e$ = 4.67×10$^{12}$ cm$^{-2}$, is about 10 times smaller than that of the pure sample, showing that Sb doping effectively reduces the bulk carriers density. For (Bi$_{1-x}$Sb$_x$)$_2$Se$_3$ with $x$ = 0.15, $R_\square$ and $\mu_H$ at RT are ~5900 $\Omega\square^{-1}$ and ~227 cm$^2$/Vs, respectively. At 10 K, $R_H$ is ~ -132 $\Omega$/T, while $R_\square$ is ~2000 $\Omega\square^{-1}$ and $\mu_H$ is ~660 cm$^2$/Vs. The reduction in mobility compared to undoped Bi$_2$Se$_3$ thin films could be due to increased scattering by Sb dopants and/or by rhombohedral/orthorhombic mixed crystalline structures. As Sb concentration increases further, not only does the resistivity $R_\square$ increase rapidly, but the temperature dependent $R_\square$ exhibits a metal-insulator transition at ~20% Sb level. For example, the $R_\square(T)$ curve for a sample with Sb concentration of ~22% is illustrated in Fig. 4C. A strongly insulating behavior with a negative slope is seen where the $R_\square$ increased from ~140 k$\Omega\square^{-1}$ at 300 K to >1 M$\Omega\square^{-1}$ at 100 K. The transport properties of selected samples are summarized in Table 1.

Table 1: Electrical transport properties of Sb-doped Bi$_2$Se$_3$ thin films.

| Sample | Average $R_H$ ($\Omega$/T) (300K) | Carrier Density (×10$^{12}$ cm$^{-2}$) (300K) | $R_\square$ (300K) | Hall Mobility (cm$^2$/Vs) at 300K | Average $R_H$ ($\Omega$/T) at low T | Carrier Density (×10$^{12}$ cm$^{-2}$) at low $T$ | $R_\square$ at low $T$ | Low $T$ Hall Mobility (cm$^2$/Vs) |
|---|---|---|---|---|---|---|---|---|
| Pure Bi$_2$Se$_3$ | -12.9 | 48.5 | 250 | 573 | -12.7 (10K) | 49.2 (10K) | 88 (10K) | 1450 (10K) |
| 15% Sb | -134 | 4.6 | 5900 | 227 | -132 (10K) | 4.7 (10K) | 2000 (10K) | 660 (10K) |
| 17% Sb | -204 | 3.1 | 13800 | 147 | -270 (30K) | 2.3 (30K) | 6500 (30K) | 415 (30K) |
| 22% Sb | -231 | 2.7 | 140000 | 17 | NA | NA | NA | NA |

As we have pointed out, at $x = 0$ when the electron density is high ($>10^{13}/cm^2$), a sample's transport is dominated by bulk carriers, similar to pure $Bi_2Se_3$ films grown by MBE. The high density of bulk carriers ($>10^{19}/cm^3$) in pure $Bi_2Se_3$ nanosheets is believed to be due to Se vacancies that overwhelm the surface electrons which normally have density on the order of $10^{12}/cm^2$.[16, 18] The magnetic field ($B$) dependent Hall resistance $R_{xy}(B)$, also shows a very weak non-linear $B$-dependence (Supporting Information, Figure S1), supporting the picture that bulk carriers dominate in band transport if one uses a two-band (bulk plus surface) theory to model the system. As the Sb doping level increases, the bulk electron density is greatly suppressed and the metallic surface states are expected to contribute more to the total conductivity of sample. This increased transport contribution from metallic surface states is manifested by two features in our data. First, as the doping density approaches the critical point where the whole sample turns insulating, a peak like behavior is observed in the $R_\square (T)$ curve due to the effect of an insulating bulk combined with a metallic surface, as shown by the kink feature ~50K for the 17% Sb doped sample in Figure 4C. This effect is similar to that observed in the bulk ternary/quaternary crystals in which the combination of insulating bulk and metallic surface gives rise to an insulating like $R(T)$ curve which saturates at low $T$.[22, 23] Moreover, in this regime of doping concentration, in contrast to what is seen in a pure sample, a clear non-linear $B$-dependence is seen in the Hall resistance data (Supporting Information, Figure S1), attesting to the increased contribution from surface states in transport.

At Sb concentrations higher than ~22%, the sample resistance became so high that it was impossible to perform electrical measurements. This is consistent with the fact that at high Sb%, the ternary $(Bi_{1-x}Sb_x)_2Se_3$ becomes a conventional band insulator[23] with a mixed rhombohedral/orthorhombic phase. The drastic change in the temperature dependent resistivity reveals a MIT at $x\sim 20\%$ (see the $x = 0.17$ and $x = 0.22$ curves in Fig. 4(c)). This general behavior of the MIT in our $(Bi_{1-x}Sb_x)_2Se_3$ nanosheets is consistent with the transition from TI to a band insulator in MBE-grown $(Bi_{1-x}In_x)_2Se_3$ thin films.[12] However, in our $(Bi_{1-x}Sb_x)_2Se_3$ nanosheets two phases (rhombohedral and orthorhombic) coexist for the $x$ value at which the MIT occurs. Since the rhombohedral phase of $(Bi_{1-x}Sb_x)_2Se_3$ is presumably topologically non-trivial, akin to $Bi_2Se_3$, but the orthorhombic phase is topologically trivial, similar to $Sb_2Se_3$, it is likely that the MIT at $x\sim 20\%$ originates from percolation of insulating orthorhombic grains through conductive background of $(Bi_{1-x}Sb_x)_2Se_3$ with rhombohedral structure.

Such a percolation picture appears to be consistent with the morphological changes of the samples (from triangular/hexagonal shapes to irregular shapes as sown in Figs. 2A and 2B). The threshold value of orthorhombic phase percentage appears to be reasonable for a percolation transition.[39] In view of the mixed phase coexistence, as evidenced by the presence of Raman modes from both the $Bi_2Se_3$ rhombohedral crystal lattice and the orthorhombic $Sb_2Se_3$ crystalline structure near the MIT ($x \sim 0.20$), we believe that the system is not transforming into a pure band insulator, but instead into a mix of the topologically trivial and non-trivial phases of $(Bi_{1-x}Sb_x)_2Se_3$ during the transition. The insets to Fig. 4C schematically illustrate this transition. For $x \leq 0.17$, the rhombohedral phase of the $Bi_2Se_3$ host lattice largely dominates, and conducting surface states (SS) exist in the bulk band gap between the conduction band (CB) and valence band (VB). For $x \sim 0.22$, the orthorhombic phase of the $Sb_2Se_3$ guest lattice contributes significantly to the physical properties of the nanosheets. This orthorhombic phase of $Sb_2Se_3$ does not have conductive SS and thus blocks the conductivity of the sample above the percolation threshold.

In summary, we have demonstrated that large areas of ternary $(Bi_{1-x}Sb_x)_2Se_3$ nanosheets with a wide range of Sb doping levels can be grown on mica *via* van der Waals epitaxy. The bulk carrier densities are suppressed in the Sb-doped $Bi_2Se_3$ nanosheets by more than an order of magnitude as Sb% increases to ~10%. A metal-insulator transition was observed near a Sb fraction of 20%. Such a transition is explained by the emergence and percolation of topologically trivial and insulating orthorhombic phase in the background of topological insulator rhombohedral phase with conductive surface states. Raman spectroscopy confirms vibrational modes from both Sb-Sb and Sb-Se bonds of $Sb_2Se_3$ at a doping level of ~20%. Our results not only demonstrate that systematic doping is a viable strategy to obtain large-area thin film topological insulators with low bulk carrier densities, but also show that $(Bi_{1-x}Sb_x)_2Se_3$ nanosheets with tunable Sb concentration can be a versatile platform for the investigation of phase transitions in TI materials.

## ACKNOWLEDGMENTS

X.P.A.G. and R.M.S. acknowledge the CWRU IAIG pilot grant for support of this research. X.P.A. G. also thanks NSF (Grant No. DMR-1151534) for partial funding support. Z.H.W. acknowledges the China Scholarship Council for a scholarship supporting her visit to CWRU. R. H. acknowledges support from UNI Faculty Summer Fellowship.

**Author Contribution Statement** C.H.Lee and Z.H. Wang performed material synthesis; R. He and C. Delaney performed Raman spectroscopy; C.H. Lee and R.L.J. Qiu performed electrical characterization; A. Kumar performed TEM characterization of sample; R. L.J. Qiu, B. Beck and T.E. Kidd performed AFM characterization of sample; C.C. Chancey and R. He, and T. E. Kidd analysed Raman spectra; R.M. Sankaran and X.P.A. Gao conceived and supervised the experiments; all authors contributed to the writing of the manuscript.

**REFERENCES**

1   L. Fu, C. L. Kane and E. J. Mele, *Phys. Rev. Lett.,* 2007, **98**, 106803.

2   H. Zhang, C.-X. Liu, X.-L. Qi, X. Dai, Z. Fang and S.-C. Zhang, *Nature Phys.,* 2009, **5**, 438-442.

3   Y. Xia, D. Qian, D. Hsieh, L. Wray, A. Pal, H. Lin, A. Bansil, D. Grauer, Y. S. Hor, R. J. Cava and M. Z. Hasan, *Nature Phys.,* 2009, **5**, 398-402.

4   Y. L. Chen, J. G. Analytis, J.-H. Chu, Z. K. Liu, S.-K. Mo, X. L. Qi, H. J. Zhang, D. H. Lu, X. Dai, Z. Fang, S. C. Zhang, I. R. Fisher, Z. Hussain and Z.-X. Shen, *Science,* 2009, **325**, 178-181.

5   J. Linder, Y. Tanaka, T. Yokoyama, A. Sudbø and N. Nagaosa, *Phys. Rev. Lett.,* 2010, **104**, 067001.

6   A. M. Essin, J. E. Moore and D. Vanderbilt, *Phys. Rev. Lett.,* 2009, **102**, 146805.

7   M. Z. Hasan and C. L. Kane, *Rev. Mod. Phys,* 2010, **82**, 3045-3067.

8   J. G. Analytis, R. D. McDonald, S. C. Riggs, J.-H. Chu, G. S. Boebinger and I. R. Fisher, *Nature Phys.,* 2010, **6**, 960-964.

9   D. Kong, J. J. Cha, K. Lai, H. Peng, J. G. Analytis, S. Meister, Y. Chen, H.-J. Zhang, I. R. Fisher, Z.-X. Shen and Y. Cui, *ACS Nano,* 2011, **5**, 4698-4703.

10  T. Sato, K. Segawa, K. Kosaka, S. Souma, K. Nakayama, K. Eto, T. Minami, Y. Ando and T. Takahashi, *Nature Phys.,* 2011, **7**, 840-844.

11  S.-Y. Xu, Y. Xia, L. A. Wray, S. Jia, F. Meier, J. H. Dil, J. Osterwalder, B. Slomski, A. Bansil, H. Lin, R. J. Cava and M. Z. Hasan, *Science,* 2011, **332**, 560-564.

12  M. Brahlek, N. Bansal, N. Koirala, S.-Y. Xu, M. Neupane, C. Liu, M. Z. Hasan and S. Oh, *Phys. Rev. Lett.,* 2012, **109**, 186403.

13  H. Peng, K. Lai, D. Kong, S. Meister, Y. Chen, X.-L. Qi, S.-C. Zhang, Z.-X. Shen and Y. Cui, *Nature Mater.,* 2010, **9**, 225-229.

14  H. Tang, D. Liang, R. L. J. Qiu and X. P. A. Gao, *ACS Nano,* 2011, **5**, 7510-7516.

15  P. Cheng, C. Song, T. Zhang, Y. Zhang, Y. Wang, J.-F. Jia, J. Wang, Y. Wang, B.-F. Zhu, X. Chen, X. Ma, K. He, L. Wang, X. Dai, Z. Fang, X. Xie, X.-L. Qi, C.-X. Liu, S.-C. Zhang and Q.-K. Xue, *Phys. Rev. Lett.,* 2010, **105**, 076801.

16  A. A. Taskin, S. Sasaki, K. Segawa and Y. Ando, *Phys. Rev. Lett.,* 2012, **109**, 066803.

17  H. D. Li, Z. Y. Wang, X. Kan, X. Guo, H. T. He, Z. Wang, J. N. Wang, T. L. Wong, N. Wang and M. H. Xie, *New J. Phys.,* 2010, **12**, 103038.

18  N. Bansal, Y. S. Kim, M. Brahlek, E. Edrey and S. Oh, *Phys. Rev. Lett.,* 2012, **109**, 116804.

19  X.-L. Qi, R. Li, J. Zang and S.-C. Zhang, *Science,* 2009, **323**, 1184-1187.

20  D. Hsieh, Y. Xia, D. Qian, L. Wray, J. H. Dil, F. Meier, J. Osterwalder, L. Patthey, J. G. Checkelsky, N. P. Ong, A. V. Fedorov, H. Lin, A. Bansil, D. Grauer, Y. S. Hor, R. J. Cava and M. Z. Hasan, *Nature,* 2009, **460**, 1101-1105.

21  S. S. Hong, J. J. Cha, D. Kong and Y. Cui, *Nature Commun.,* 2012, **3**, 757.

22  Z. Ren, A. A. Taskin, S. Sasaki, K. Segawa and Y. Ando, *Phys. Rev. B,* 2011, **84**, 075316.

23  Z. Ren, A. A. Taskin, S. Sasaki, K. Segawa and Y. Ando, *Phys. Rev. B,* 2011, **84**, 165311.

24  C. Niu, Y. Dai, Y. Zhu, Y. Ma, L. Yu, S. Han and B. Huang, *Sci. Rep.,* 2012, **2**, 976.

25  J. Chen, H. J. Qin, F. Yang, J. Liu, T. Guan, F. M. Qu, G. H. Zhang, J. R. Shi, X. C. Xie, C. L. Yang, K. H. Wu, Y. Q. Li and L. Lu, *Phys. Rev. Lett.,* 2010, **105**, 176602.

26  J. G. Checkelsky, Y. S. Hor, R. J. Cava and N. P. Ong, *Phys. Rev. Lett.,* 2011, **106**, 196801.

27  F. Xiu, L. He, Y. Wang, L. Cheng, L.-T. Chang, M. Lang, G. Huang, X. Kou, Y. Zhou, X. Jiang, Z. Chen, J. Zou, A. Shailos and K. L. Wang, *Nature Nanotech,* 2011, **6**, 216-221.

28  Y. Wang, F. Xiu, L. Cheng, L. He, M. Lang, J. Tang, X. Kou, X. Yu, X. Jiang, Z. Chen, J. Zou and K. L. Wang, *Nano Letters,* 2012, **12**, 1170-1175.

29  J. Zhang, C.-Z. Chang, Z. Zhang, J. Wen, X. Feng, K. Li, M. Liu, K. He, L. Wang, X. Chen, Q.-K. Xue, X. Ma and Y. Wang, *Nature Commun.,* 2011, **2**, 574.

30  H. Peng, W. Dang, J. Cao, Y. Chen, D. Wu, W. Zheng, H. Li, Z.-X. Shen and Z. Liu, *Nature Chem.,* 2012, **4**, 281-286.

31  R. Vadapoo, S. Krishnan, H. Yilmaz and C. Marin, *Phys. Status Solidi B,* 2011, **248**, 700-705.

32    I. Teramoto and S. Takayanagi, *J. Phys. Chem. Solids,* 1961, **19**, 124-129.

33    V. V. Sobolev, S. D. Shutov, Y. V. Popov and S. N. Shestatskii, *Phys. Status Solidi B,* 1968, **30**, 349-355.

34    Z. Wei, Y. Rui, Z. Hai-Jun, D. Xi and F. Zhong, *New J. Phys.,* 2010, **12**, 065013.

35    R. W. G. Wyckoff, *Crystal Structures* (John Wiley and Sons, New York, 1964).

36    W. Richter and C. R. Becker, *Phys. Status Solidi B,* 1977, **84**, 619-628.

37    J. Zhang, Z. Peng, A. Soni, Y. Zhao, Y. Xiong, B. Peng, J. Wang, M. S. Dresselhaus and Q. Xiong, *Nano Letters,* 2011, **11**, 2407-2414.

38    Z. G. Ivanova, E. Cernoskova, V. S. Vassilev and S. V. Boycheva, *Mater. Lett.,* 2003, **57**, 1025-1028.

39    S. Kirkpatrick, *Rev. Mod. Phys,* 1973, **45**, 574-588.

# Metal-Insulator Transition in Variably Doped (Bi$_{1-x}$Sb$_x$)$_2$Se$_3$ Nanosheets

Chee Huei Lee,[†,‡] Rui He,[§,*] ZhenHua Wang,[†,¶] Richard L.J. Qiu,[†] Ajay Kumar,[‡] Conor Delaney,[§] Ben Beck,[§] T. E. Kidd,[§] C. C. Chancey,[§] R. Mohan Sankaran,[‡] and Xuan P. A. Gao,[†,*]

[†]Department of Physics, Case Western Reserve University, Cleveland, OH 44106, U.S.A.

[‡]Department of Chemical Engineering, Case Western Reserve University, Cleveland, OH 44106, U.S.A.

[§]Department of Physics, University of Northern Iowa, Cedar Falls, IA 50614, U.S.A.

[¶]Shenyang National Laboratory for Materials Science, Institute of Metal Research, and International Centre for Materials Physics, Chinese Academy of Sciences, Shenyang 110016, People's Republic of China

*Email: (R.H.) rui.he@uni.edu; (X.P.A.G.) xuan.gao@case.edu

**Evidence of multi-band transport in doped (Bi$_{1-x}$Sb$_x$)$_2$Se$_3$ nanosheets in Hall effect data**

In un-doped Bi$_2$Se$_3$ nanosheet, due the high bulk carrier density ($>10^{19}$/cm$^3$) from Se vacancies, the transport is dominated by the bulk carriers and Hall resistance shows a nearly perfect linear relationship with perpendicular magnetic field, $B$ (solid black line, Figure S1). The extracted Hall slope, $R_H = R_{xy}/B$, from linear fitting of $R_{xy}(B)$ is expected to relate to the bulk carrier density $n_b = -1/(eR_H)$. However, as Sb doping reduces the bulk carrier density and conductivity contribution, the effect of surface states conduction in topological insulators is expected to be readily observable through non-linear Hall slope data.[1,2] In a two band transport system, the $R_{xy}$ is given by [2]

$$R_{xy}(B) = -(B/e)[(n_s\mu_s^2 + n_b\mu_b^2) + B^2\mu_s^2\mu_b^2(n_s+n_b)] / [(n_s\mu_s + n_b\mu_b)^2 + B^2\mu_s^2\mu_b^2(n_s+n_b)^2] \quad (1)$$

where $n_s$, $\mu_s$ and $n_b$, $\mu_b$ represent the carrier density and mobility for the surface and bulk channel. Eq. 1 approaches the asymptotic behavior $R_{xy}/B(\infty) = -(1/e)(n_s+n_b)^{-1}$ in the strong field limit, and approaches $R_{xy}/B(0) = -(1/e)(n_s\mu_s^2 + n_b\mu_b^2)/(n_s\mu_s + n_b\mu_b)^2$ in the small field limit. If $\mu_b \neq \mu_s$, then one has a field dependent $R_{xy}/B$ and the field Hall slope is larger than the high field slope: $|R_{xy}/B(0)| > R_{xy}/B(\infty)$. This effect was indeed observed in our doped $(Bi_{1-x}Sb_x)_2Se_3$ nanosheet samples. Figure S1 A compares the $R_{xy}$ data for a pure sample and a $(Bi_{1-x}Sb_x)_2Se_3$ nanosheet sample with $x\sim0.17$. The non-linear $R_{xy}(B)$ data in doped sample is illustrated by the linear extrapolations of high field data not passing origin (dashed lines). The difference in Hall effect between un-doped and doped samples is further highlighted in plotting $R_{xy}/B$ vs. in Figure S1B. In contrast to the constant Hall slope for un-doped sample, the magnitude of $R_{xy}/B$ gradually decreases and approaches a constant value at high $B$, as Eq.1 predicts.

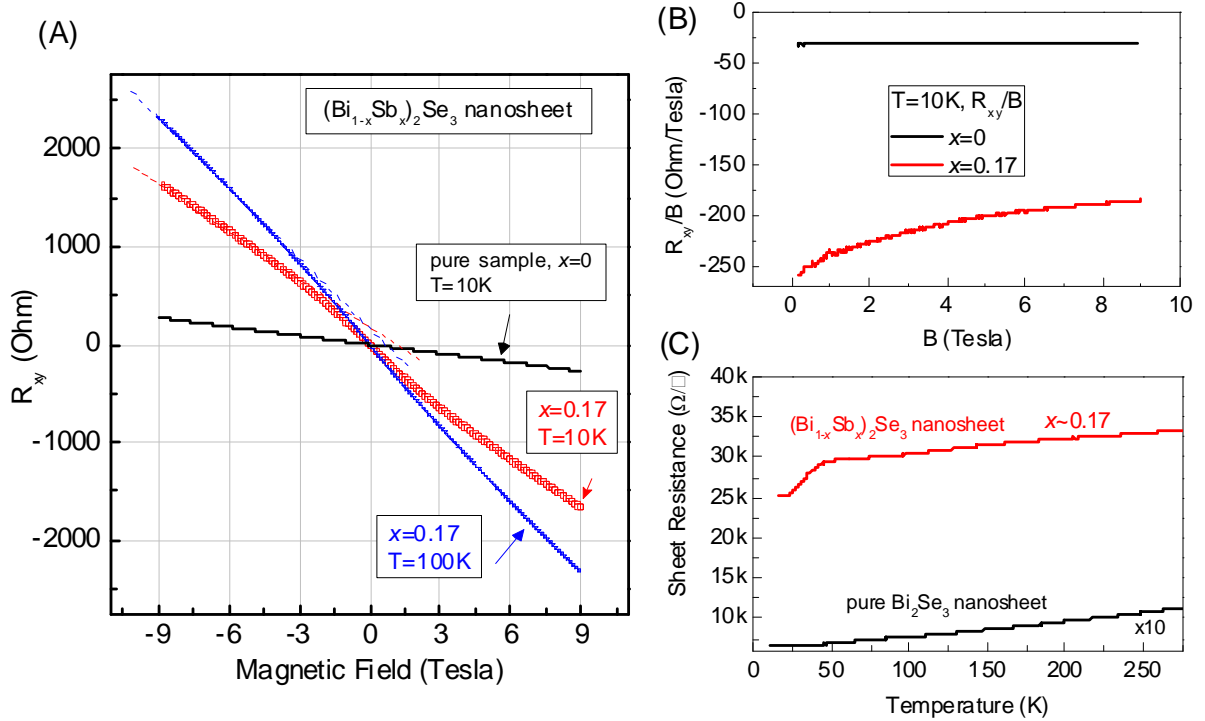

**Fig. S1.** (A) Field dependent Hall resistance for pure $Bi_2Se_3$ nanosheet and $(Bi_{1-x}Sb_x)_2Se_3$ nanosheet with $x=0.17$ at various temperatures, to show the nonlinear $R_{xy}$ vs $B$ in doped sample. (B) The Hall slope $R_{xy}/B$ plotted as a function of $B$ for pure and doped samples, highlighting the constant slope in pure sample vs. a $B$-dependent slope in doped sample. The non-linear $R_{xy}$ vs $B$ in doped sample is attributed to the increased contribution from the surface states channel in a two band (bulk plus surface) transport system. (C) Sheet resistance vs. temperature for the two $(Bi_{1-x}Sb_x)_2Se_3$ nanosheet samples in (A) and (B).

## REFERENCES

(1) A. A. Taskin, S. Sasaki, K. Segawa, Y. Ando, *Phys. Rev. Lett.* 2012, 109, 066803.

(2) N. Bansal, Y. S. Kim, M. Brahlek, E. Edrey, S. Oh, *Phys. Rev. Lett.* 2012, 109, 116804.